\documentclass{article}

\usepackage[preprint]{neurips_2024}

\usepackage[utf8]{inputenc} 
\usepackage[T1]{fontenc}    
\usepackage{hyperref}       
\usepackage{url}            
\usepackage{booktabs}       
\usepackage{amsfonts}       
\usepackage{nicefrac}       
\usepackage{microtype}      
\usepackage{xcolor}         
\usepackage{amsmath,amssymb,amsfonts}
\usepackage{algorithmic}
\usepackage{graphicx}
\usepackage{epstopdf}
\usepackage{textcomp}
\usepackage{tikz}
\usetikzlibrary{positioning, shapes.geometric}
\usepackage{xcolor}
\usepackage[linesnumbered,ruled,vlined]{algorithm2e}
\usepackage{amsthm} 
\usepackage{mathrsfs}
\newtheorem{theorem}{Theorem}
\newtheorem{corollary}{Corollary}
\newtheorem{lemma}{Lemma}
\newtheorem{definition}{Definition}
\newtheorem{assumption}{Assumption}

\title{Smart Predict-then-Optimize Method with Dependent Data: Risk Bounds and Calibration of Autoregression}

\author{
  Jixian Liu\\
  Xi'an Jiaotong University\\
  Xi'an, Shaanxi 710049 \\
  \texttt{2204110794@stu.xjtu.edu.cn} \\
  \And  
  Tao Xu\\
  Shanghai Jiaotong University\\
  Shanghai, Shanghai 200240\\
  \texttt{Zerken@sjtu.edu.cn} \\
  \And
  Jianping He\\
  Shanghai Jiaotong University\\
  Shanghai, Shanghai 200240\\
  \texttt{jphe@sjtu.edu.cn}
  \And
  Chongrong Fang\\
  Shanghai Jiaotong University\\
  Shanghai, Shanghai 200240\\
  \texttt{crfang@sjtu.edu.cn@sjtu.edu.cn}
}

\definecolor{customblue}{RGB}{0, 100, 204}
\definecolor{customgreen}{RGB}{0, 180, 204}
\definecolor{customorange}{RGB}{255, 128, 0}

\tikzstyle{startstop} = [rectangle, rounded corners, minimum width=3cm, minimum height=1cm, text centered, draw=white, fill=customblue]
\tikzstyle{process} = [rectangle, minimum width=3cm, minimum height=1cm, text centered, draw=wight, fill=customgreen]
\tikzstyle{decision} = [diamond, minimum width=3cm, minimum height=1cm, text centered, draw=white, fill=green]
\tikzstyle{arrow} = [thick,->,>=stealth]

\begin{document}
\maketitle
\begin{abstract}
The predict-then-optimize (PTO) framework is indispensable for addressing practical stochastic decision-making tasks. It consists of two crucial steps: initially predicting unknown parameters of an optimization model and subsequently solving the problem based on these predictions. Elmachtoub and Grigas \cite{elmachtoub2022smart} introduced the Smart Predict-then-Optimize (SPO) loss for the framework, which gauges the decision error arising from predicted parameters, and a convex surrogate, the SPO+ loss, which incorporates the underlying structure of the optimization model. The consistency of these different loss functions is guaranteed under the assumption of i.i.d. training data. Nevertheless, various types of data are often dependent, such as power load fluctuations over time. This dependent nature can lead to diminished model performance in testing or real-world applications. Motivated to make intelligent predictions for time series data, we present an autoregressive SPO method directly targeting the optimization problem at the decision stage in this paper, where the conditions of consistency are no longer met. Therefore, we first analyze the generalization bounds of the SPO loss within our autoregressive model. Subsequently, the uniform calibration results in Liu and Grigas \cite{liu2021risk} are extended in the proposed model. Finally, we conduct experiments to empirically demonstrate the effectiveness of the SPO+ surrogate compared to the absolute loss and the least squares loss, especially when the cost vectors are determined by stationary dynamical systems and demonstrate the relationship between normalized regret and mixing coefficients.
\end{abstract}

\section{\scshape Introduction}\label{Sec. 1 Intro}
The ability to make wise decisions in machine learning is becoming an increasingly important research hotspot. In many practical scenarios, such as portfolio allocation and multi-class classification problems, the predict-then-optimize framework is commonly used to aid decision-making. This framework involves first predicting the unknown parameters of an optimization model and then solving the optimization problem based on these predicted parameters using contextual information, such as weather data, historical user preferences, and other relevant factors. Nevertheless, prediction accuracy often diverges from decision accuracy. As a result, individuals frequently prioritize optimizing the final goal over making precise predictions of uncertain contextual variables. Consequently, substantial research efforts focus on developing efficient algorithms that directly optimize the ultimate objective of the problem 
\cite{elmachtoub2022smart,wilder2019melding,poganvcic2019differentiation, amos2017optnet, donti2017task,bertsimas2020predictive}.

Particularly, for the case where the optimization problem has a linear objective, Elmachtoub and Grigas \cite{elmachtoub2022smart} developed a framework called Smart "predict then optimize"(SPO), which designed the SPO loss function which directly measures the regret between the cost generated by decision and the optimal cost and the convex surrogate loss function for the SPO loss which is called SPO+ loss and has a Fisher consistency property with the SPO loss. Recent works have studied a lot of statistical properties of the SPO loss and SPO+ loss. El Balghiti et al \cite{el2019generalization} analyzed an $O\left(\frac{\log(C'n)}{\sqrt{n}}\right)$ generalization bound of the SPO loss between the empirical error and the generalization one, where $C'$ is a constant and $n$ is the training data size. The generalization and regret convergence rate is analyzed by Hu et al \cite{hu2022fast}. Nevertheless, Fisher consistency often can not be obtained due to the limited data in practice because it needs full information on the underlying distribution, so Liu and Grigas \cite{liu2021risk} develop risk bounds that allow one to translate an approximate guarantee on the risk of a surrogate loss function to the guarantee on the SPO risk, that is calibrated the tolerance of the surrogate loss and the excess SPO risk. 

Despite the limited data, there is another essential factor that the data may be dependent in many real-world applications that can degenerate the performance of the prediction model. Many current works focus on training from dependent data by reducing the formulated problem to independent learning via blocking techniques. In this work, the scenarios when the contextual data becomes dependent are investigated, such as load forecasting by simply using the historical load\cite{hong2020energy} and the Fisher consistency can not be guaranteed even with abundant data due to the dependent property of the stationary $\beta$-mixing process. A stationary mixing sequence is often used in modeling these scenarios, which implies a dependence between observations that diminishes with time \cite{kuznetsov2020discrepancy,mohri2008rademacher,kuznetsov2018theory,lozano2005convergence,yu1994rates,mohri2007stability}. We leverage the independent block techniques in \cite{yu1994rates} to derive similar calibration bound for the stationary mixing processes comparing to the i.i.d. setting. To be specific, We investigate a similar setup like the "Smart Predict-them-Optimize" in the context of offline contextual optimization problems \cite{elmachtoub2022smart} to highlight the central idea. The main contributions are organized as follows.

\begin{itemize} 
    \item We analyze the generalization bound of the SPO loss for a stationary and $\beta$-mixing process and derive the uniform calibration between the SPO+ loss and the SPO loss for the stationary and $\beta$-mixing process.
    \item We design an autoregression model with the SPO loss to directly solve the formulated optimization problem rather than using the traditional PTO framework.
    \item We perform experiments on the formulated optimization problem whose cost vectors are automatically generated from a stochastic dynamical system with zero-bias noise by using the SPO+ loss function to compare with conventional mean absolute error and least square prediction error losses, focusing on different dynamical systems. 
\end{itemize}

The rest of the paper is organized as follows. Sec \ref{Preliminaries} defines the problem and gives preliminaries of the SPO Framework for the stationary mixing processes. Sec \ref{Proposed Algorithm} and \ref{Generalization} present the corresponding autoregressive algorithm with a fixed memory and analyze the generalization bound of SPO loss and the uniform calibration between the SPO+ loss and SPO loss using dependent data. Sec \ref{Computational} shows the numerical experiments. Finally, Sec \ref{Conclusions} concludes this paper and discusses the possible future directions.

\textbf{Notation}. Let $I_{p}$ denote the $p \times p$ identity matrix for any positive integer $p$. For $\overline{y} \in \mathbb{R}^{d}$ and a positive semi-definite matrix $\Sigma \in \mathbb{R}^{d \times d}$, let $\mathcal{N}(\overline{y}, \Sigma)$ denote the normal distribution with mean $\overline{y}$ and covariance matrix $\Sigma$, typically denoted as $\mathcal{N}(\overline{y}, \Sigma) = \frac{1}{\sqrt{(2\pi)^d |\Sigma|}} e^{-\frac{1}{2} (\mathbf{y} - \overline{y})^\top \Sigma^{-1} (\mathbf{y} - \overline{y})}$ and $\mathcal{P}_{Gaussian}$ is the Gaussian distribution class. $\mathfrak{X}, \mathfrak{Y}$ is the feasible region of state vectors and cost vectors, respectively. $S$ is the feasible region of the decision vector $w$. The term $l$ is the lag length of the autoregression model. $\omega_{S}(y)$ is the "linear optimization gap" of $S$ \cite{el2019generalization}. $\omega_{S}(\mathfrak{Y}) := \sup_{y \in \mathfrak{Y}}(\omega_{S}(y))$.

\section{\scshape Preliminaries and Problem Formulation}\label{Preliminaries}
\subsection{``Smart Predict, then Optimize" Framework}
Firstly, we describe the "Predict, then Optimize" framework which is critical to many practical optimization problems. Specifically, for a nominal optimization problem of interest with a linear objective, where the decision variable $w \in \mathbb{R}^{d}$ and $w$’s feasible region $S \subseteq \mathbb{R}^{d}$ are well-defined with certainty, its cost vector $y \in \mathbb{R}^{d}$ is commonly not known when the decision should be made.

However, we can get the related feature vector $x \in \mathbb{R}^{p}$ of the optimization problem, where $(x,y)$ are random variables subject to some joint distribution. Let $\mathcal{D}_x$ be the conditional distribution of $y$ given $x$. Then the goal of the decision maker is to solve the contextual stochastic optimization problem \cite{elmachtoub2022smart}:  
\begin{equation}
    \textbf{Problem~1:} ~\min\limits_{w \in S}\mathbb{E}_{\mathbf{y} \sim \mathcal{D}_x} [\mathbf{y}^{\mathrm{T}}w|x]=\min\limits_{w \in S}\mathbb{E}_{\mathbf{y}\sim \mathcal{D}_x} [\mathbf{y}^{\mathrm{T}}|x]w 
\label{pro. 1}
\end{equation}

To derive a model for predicting cost vectors, denoted as a cost vector predictor function $f: \mathfrak{X} \rightarrow \mathbb{R}^{d}$, we can utilize machine learning techniques to learn the underlying distribution from observed data $\left\{(x_1,y_1),\ldots, (x_n, y_n)\right\}$. These data points are assumed to be i.i.d. sampled from a distribution. In the traditional PTO framework, we first get predicted cost vector $\hat{y}$ then solve the optimization problem $\min_{w \in S}\hat{y}^{\mathrm{T}}w$

Nevertheless, it is often frustrating that we can hardly get the full knowledge of the above distributions in most scenarios and are only access to a limited batch of data $\{(x_i,y_i)\}, i = 1, 2, \ldots, N$, which means that the prediction $\hat{y}$ generated by the obtained data and contextual feature is almost different from the expectation $\mathbb{E}_{\mathbf{y} \sim \mathcal{D}_x}[\mathbf{y}^{T}|x]w$. To overcome the misalignment between the prediction accuracy and the optimization results. A "Smart predict, then optimize" framework is designed compared with the traditional PTO methods\cite{wang2006cope, mukhopadhyay2017prioritized}, which bridged the misalignment by introducing the Smart "Predict then Optimize" loss function which directly designed to be oriented to shrinking the decision error in the optimization stage:
\begin{equation}
    \ell_{SPO} := y^{\mathrm{T}}w^{*}(\hat{y}) - y^{\mathrm{T}}w^{*}(y),
\end{equation}
where $\hat{y} \in \mathbb{R}^{d}$ is the predicted cost vector and $y \in \mathbb{R}^{d}$ is the realized cost vector.

The convex surrogate SPO loss: SPO+ loss is also introduced in \cite{elmachtoub2022smart} due to the non-convexity and even discontinuous property of the SPO loss, as:
\begin{equation}
    \ell_{SPO+}(\hat{y},y) := \max_{w\in S}{(y-2\hat{y})^{\mathrm{T}w}} + 2\hat{y}^{\mathrm{T}}w^{*}(y) - c^{\mathrm{T}}w^{*}(y).
\end{equation}

The SPO+ loss retains its consideration for the downstream optimization problem \ref{pro. 1} and the structure of the feasible region S, unlike loss functions that solely focus on prediction error. As highlighted by \cite{elmachtoub2022smart}, efficient optimization of the SPO+ loss can be achieved through linear/conic optimization reformulations and via (stochastic) gradient descent methods, making it suitable for large datasets. The SPO loss offered theoretical and empirical validation for the SPO+ loss function, including its derivation via duality theory, promising experimental outcomes on instances of shortest path and portfolio optimization, and the following theorem establishing the Fisher consistency between the SPO loss and the surrogate one. 

\subsection{Uniform Calibration}
According to the consistency results, as the continuous and the symmetric conditions hold for the conditional distribution $\mathbb{P}(\cdot|x)$, $\overline{y}:= \mathbb{E}_{y \sim \mathbb{P}(\cdot|x)}$ is the minimizer of the loss functions for the nominal problem. Nevertheless, due to the limited data, the fisher consistency among different loss functions cannot be applicable. Although the consistency results may not apply to the limited data scenario,  It is still worthwhile to characterize the relationship among the generalization bounds of different loss functions, such as the least squares loss function $\ell_{2}$ \cite{ho2022risk}, the  SPO+ loss function \cite{liu2021risk} and SPO loss function.

\begin{definition}($\ell_{SPO}$-calibrated \cite{liu2021risk}) For a given surrogate loss function $\ell$, we say $\ell$ is $\ell_{SPO}$-calibrated with respect to distribution $\mathbb{P}$ if there exists a function $\delta_{\ell}(\cdot): \mathbb{R}_{+} \rightarrow \mathbb{R}_{+}$ such that for all $x \in \mathfrak{X}$, $y \in \mathfrak{Y}$ and $\epsilon > 0$, it holds that
\begin{equation}
\label{eq. cali}
    \mathbb{E}[\ell(\hat{y}, y) \mid x]-\inf _{y^{\prime}} \mathbb{E}\left[\ell\left(y^{\prime}, y\right) \mid x\right]<\delta_{\ell}(\epsilon) \Rightarrow \mathbb{E}\left[\ell_{\mathrm{SPO}}(\hat{y}, y) \mid x\right]-\inf _{y^{\prime}} \mathbb{E}\left[\ell_{\mathrm{SPO}}\left(y^{\prime}, y\right) \mid x\right]<\epsilon,
\end{equation}
where $\hat{y} \in \mathfrak{Y}$ is the predicted value of a predtion model $\hat{f}$. Besides, if it holds for all $\mathbb{P} \in \mathcal{P}$, where $\mathcal{P}$ is a class of distributions on $\mathfrak{X} \times \mathfrak{Y}$, then we say that $\ell$ is uniformly calibrated concerning the class of distributions $\mathcal{P}$.
\label{def. calibration}
\end{definition}

In this paper, we derive the excess SPO risk between the optimal predictor $\hat{f}_{SPO+}^{n}$ with $n$ training data that minimize the empirical SPO+ loss and the true optimal predictor in different feasible regions using the calibration function $\delta_{\ell}$. 

\subsection{Stationary $\beta$-mixing Processes}
Despite the restriction of the limited data, the data collected from the real world is often non-i.i.d. and shows dependency chronologically, which will also end in the fact that consistency cannot be guaranteed. In this paper, we will explore the scenarios where data is not independent, rooted in stationary $\beta$-mixing processes.

\begin{definition}
(Stationary). A sequence of random variables $\mathcal{Y} = \{y_{t}\}_{t=-\infty}^{+\infty}$ is said to be stationary if for any t and non-negative integers $n$ and $k$, the block of random vectors $\{y_{t:t+n}\}$, and $\{y_{t+k:t+n+k}\}$ have the same distribution.
\end{definition}

\begin{definition}
($\beta-mixing$). Let $\mathcal{Y}$ be a stationary sequence. For any integer $i,j \in \mathbb{Z} \cup \{-\infty, +\infty\}$, $\sigma_{i}^{j}$ is the $\sigma$-algebra defined on the blocks of random variables ${y_{i:j}}$. For positive integer $k$, the $\beta$-mixing coefficients is defined as 
\begin{equation}
    \beta(k) = \sup_{n} \mathbb{E}_{B \in \sigma_{-\infty}^{n}}\left[\sup_{A\in \sigma_{n+k}^{+\infty}}|\mathbb{P}\left(A|B\right)-\mathbb{P}(A)|\right].
\end{equation}
A stochastic process is regular, or $\beta$-mixing if $\beta(k) \rightarrow 0$ as $k \rightarrow 0$.
\end{definition}

\begin{assumption}
\label{ass. 1}
    $\{y_{i}\}_{i=1}^{+\infty}$ is a stationary and $\beta$-mixing process with mixing coefficients $\beta(a), \forall a > 0$ 
\end{assumption}

Thus, a sequence of random variables is mixed when the dependence of an event on those occurring $k$ units of time in the past weakens as a function of $k$.

\subsection{Problem Formulation}

Compared with the optimization \textbf{Problem \ref{pro. 1}}, we want to extend it to an online nominal optimization problem with an observed $N$-step trajectory $ \{y\}_{i=1}^{N} \subseteq \mathbb{R}^{d} $ at the $N+1$-th step where the Assumption \ref{ass. 1} holds. $y_{k} \in \mathbb{R}^{p}$ is the cost vector and $\{y_{i}\}_{i=1}^{+\infty}$ is also a stationary and $\beta$-mixing process.

Then the online optimization problem can be formulated as follows
\begin{equation}
\begin{aligned}
\textbf{Problem~2:}
    \min\limits_{w\in S}\mathbb{E}_{\mathbf{y}_{N+1}\sim\mathcal{P}_{\mathbf{y}_{N+1}|\mathbf{y}_{1:N}}}[\mathbf{y}_{N}^\mathrm{T} w|y_{1:N}] =\min\limits_{w\in S}\mathbb{E}_{\mathbf{y}_{N+1}\sim\mathcal{P}_{\mathbf{y}_{N+1}|\mathbf{y}_{1:N}}}[\mathbf{y}_{N+1}|y_{1:N}]^\top w.
\end{aligned}
 \label{eq-2}
\end{equation}

In our problem setting, the state vectors $\{y_{i}\}_{i=1}^{N}$ for the initial $N$ steps is inaccessible. Consequently, we aim to design a prediction algorithm to directly facilitate decision-making by minimizing the empirical risk of a given loss function, which is slightly different from the empirical risk in \cite{elmachtoub2022smart} and will be defined later, then solve the Equation \ref{eq-2}. In other words, despite the possibility of a considerable margin of error between the predicted state vector and the actual one, the predictive model enables decision-makers to make more informed decisions at the $N+1$-th step when the cost vector $\mathbf{y}_{N+1}$ is unknown. 

\section{\scshape Proposed Algorithm}\label{Proposed Algorithm}
\textbf{Autoregressive Model with Fixed Memory}. To predict the observed trajectory for an unknown stochastic dynamical system $\Phi$, an intuitive idea is to use the auto-regression model to make the prediction. In the paper, we assume that the order of the autoregression model is $l \in \mathbb{N}_{+}$. For $ l+1 \leq k \leq N$, our goal is to design proper matrices $M_1, M_2, ..., M_l$ to approximate the observation value at time $k$ by using the former $l$ steps' observation value:

\begin{equation}
    \label{eq.5}
    y_{k} \approx \sum_{i=1}^{l} M_i y_{k-i} 
\end{equation}

\begin{figure}[htbp]
    \centering
    \includegraphics[width=0.68\columnwidth]{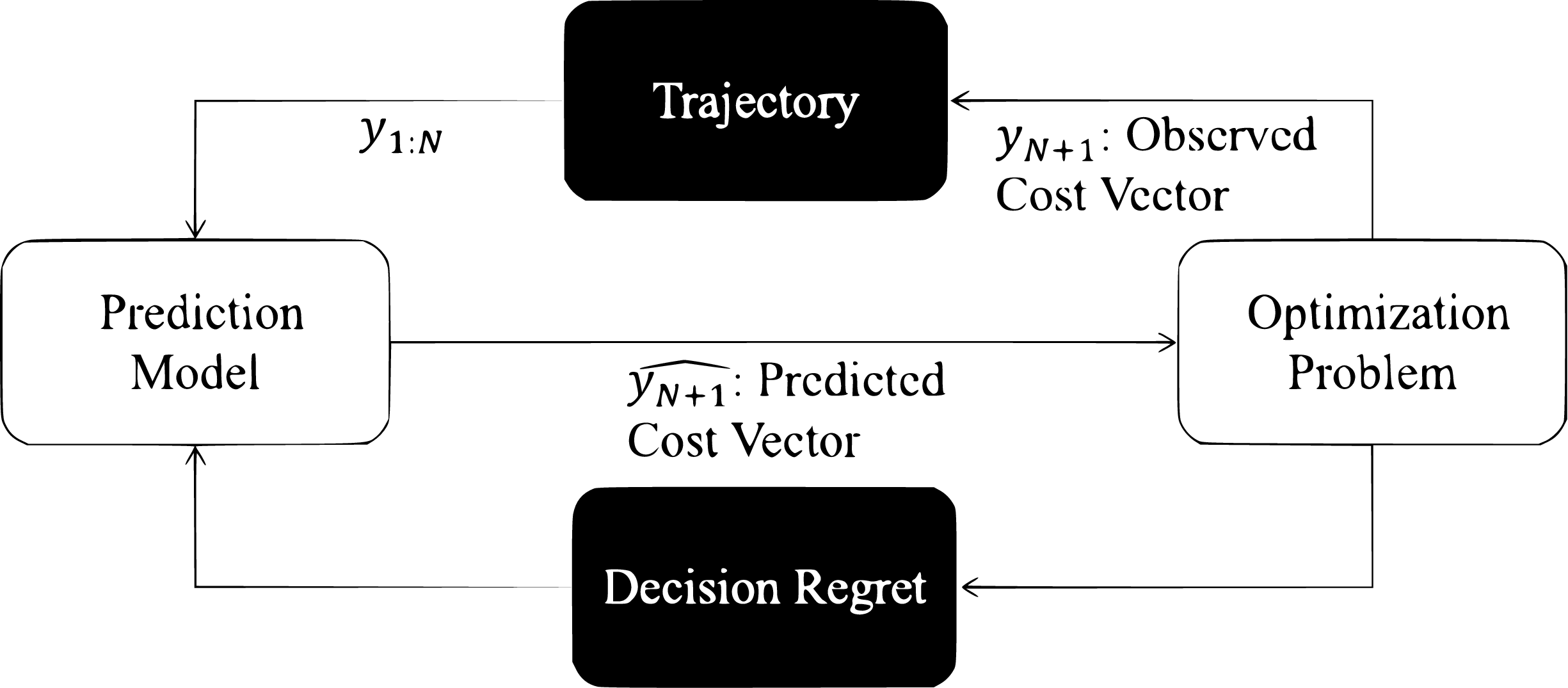}
    \caption{SPO-based Autoregression Model}
    \label{fig:enter-label}
    \label{fig: 2}
\end{figure}

\textbf{SPO-based Autoregressive Method}. Being similar to the SPO framework in the i.i.d. setting, we combined SPO loss and SPO+ loss with the autoregressive model as shown in Figure \ref{fig: 2} for the time series prediction.

\begin{definition}
\label{def: 1}
    Given a predicted cost vector at N+1 time step as $\hat{y}_{N+1}$, and the realized cost vector at N+1 time step as $y_{N+1}$ 
\end{definition}

Comparing simply minimizing the autoregressive model's prediction error, such as least squares error and absolute error, by employing methods like regression, we design a more effective method that is based on the estimated regret of the desired decision by adopting the SPO and SPO+ loss into the autoregression processes. Essentially, SPO is tailored to conduct prediction focusing on the final optimization outcome. Specifically, for our port selection issue, this results in the formulation of the empirical loss function which will be reformulated later with a slightly different from traditional ones.

\begin{algorithm}
	\renewcommand{\algorithmicrequire}{\textbf{Input:}}
	\renewcommand{\algorithmicensure}{\textbf{Output:}}
	\caption{Stochasitic Subgradient Descent for SPO-Based Autoregression with Fixed Memory}
	\label{alg. autore}
	\begin{algorithmic}[1]
		\STATE Initialization:$l,M^{1}=[M_1^{1}, M_2^{1}, M_3^{1}, ..., M_l^{1}] \in \mathbb{R}^{d \times ld}$ randomly.
		\STATE Partial Autocorrelation Function is used to determine l.
		\STATE Set i=l, t=1 and the variance tolorance $\delta_{t}$ 
		\REPEAT 
			\WHILE{$i \leq N-1$}
				\STATE Compute $\tilde{w}_i^{t} = w^{*}(2 M_t [y_{i}^{\mathrm{T}}, y_{i-1}^{\mathrm{T}}, ..., y_{i-l+1}^{\mathrm{T}}]^{\mathrm{T}}-y_{i+1})$
				\STATE Set $\tilde{G}_{i}^{t} = (w^{*}(y_{i+1})-\tilde{w}_{i}^{t})[y_{i}^{\mathrm{T}}, y_{i-1}^{\mathrm{T}}, ..., y_{i-l+1}^{\mathrm{T}}]$
				\STATE i=i+1
			\ENDWHILE
		\STATE Compute $\tilde{G^{t}} = \frac{1}{N-l} \sum_{i=1}^{N-l} \tilde{G}_{i}^{t}$
		\STATE Update $M_{t+1} = M_{t} - \alpha \tilde{G^{t}}$
		\STATE $t \leftarrow t + 1$
		\UNTIL $\Vert M_{t+1} - M_{t}\Vert \leq \delta_{t}$
	\end{algorithmic}  
\end{algorithm}

Besides, because for a fixed realized cost vector $y_{i}, i \in \mathbb{N}_{+}$, given any $\hat{y}_{i}, i \in \mathbb{N}_{+}$, $2(w^{*}(y_{i})-w^{*}(2\hat{y_{i}}-y_{i})$ is a subgredient of $\ell_{SPO+}$ at $\hat{y}_{i}$ \cite{elmachtoub2022smart}, Algorithm.\ref{alg. autore} is the stochastic subgradient descent process of the autoregressive model to get the predicted matrix $\hat{M}$ that minimizes the SPO+ loss.

\section{\scshape Generalization Bounds and Calibration Results}\label{Generalization}
As Assumption \ref{ass. 1} holds, we can now state our bounds and calibration results for the end-to-end time series forecasting. 

\textbf{Independent Blocks.} As the previous works for the analysis of the dependent data \cite{mohri2008rademacher,yu1994rates,bernstein1927extension} to derive the corresponding Generalization Bounds for the stationary and $\beta$-mixing sequence, we use the block techniques to transfer the original problem of dependent points to one that is based on independent blocks. To be specific, we split a sequence $\mathcal{Y} = \{y_{1}, ..., y_{n}\}$ into two subsequences $\mathcal{Y}_{0}$ and $\mathcal{Y}_{1}$, each of which has $m$ sequences and each sequence of them has $a$ elements. That is $n = 2a m + l$, $\mathcal{Y}_{0}$ and $\mathcal{Y}_{1}$ are defined as follows:

\begin{equation}
\begin{array}{ll}
\mathcal{Y}_{0}=\left \{ Y_{1}, Y_{2}, \ldots, Y_{m}\right \}, & \text { where } Y_{i}=\left \{ y_{(2 i-1)+1+l}, \ldots, y_{(2 i-1)+a+l}\right \}, \\
\mathcal{Y}_{1}=\left \{ Y_{1}^{(1)}, Y_{2}^{(1)}, \ldots, Y_{m}^{(1)}\right \}, & \text { where } Y_{i}^{(1)}=\left \{ y_{2 i+1+l}, \ldots, y_{2 i+a+l}\right \}.
\end{array}
\end{equation}

And we use $\mathcal{Y}_{0}$ for our generalization bounds analysis.

\textbf{Rademacher Complexity}. Being a little different from the notion of Rademacher complexity used in \cite{el2019generalization, mohri2008rademacher}, Given an observed sequence $\mathcal{Y}$, we define the empirical risk concerning the SPO loss of a predict function $f \in \mathcal{H}$ as 
\begin{equation}
    \widehat{R}_{SPO, n} = \frac{1}{n-l} \sum_{i=l+1}^{n} \ell_{\mathrm{SPO}}\left(f\left(y_{1:i-1}\right), y_{i}\right),
\end{equation}
In the fixed memory setting, we have that $f(y_{1:i-1}) = f(y_{i-l}:y_{i-1})$. We define $y_{1:0} = \varnothing$. Besides, we define the generalization loss for the stationary mixing processes as 
\begin{equation}
    R_{SPO, n}(f) = \mathbb{E}_{(y_{1:n-1},y_{n}) \sim \mathcal{D}}[\ell_{SPO}(f(y_{1:n-1}), y_{n})],
\end{equation}
where $\mathcal{D}$ is the joint distribution of $\mathcal{Y}$. Then we define the empirical Rademacher complexity of $\mathcal{H}$ concerning the SPO loss as 
\begin{equation}
    \widehat{\mathfrak{R}}_{\mathrm{SPO},\mathcal{Y}}(\mathcal{H}):=\mathbb{E}_{\sigma}\left[\sup _{f \in \mathcal{H}} \frac{1}{n-l} \sum_{i=l+1}^{n} \sigma_{i} \ell_{\mathrm{SPO}}\left(f\left(y_{1:i-1}\right), y_{i}\right)\right],
\end{equation}
where $\sigma_{i}, i=1,2,\ldots,n,$ are i.i.d. Rademacher random variables. $\mathfrak{R}^{n}_{SPO}(\mathcal{H}) = \mathbb{E}_{(y_{1:n-1.y_{n}})\sim\mathcal{D}}[\widehat{\mathfrak{R}}_{\mathrm{SPO},\mathcal{Y}}(\mathcal{H})]$ is the expected Rademacher complexity. For rare cases where a distribution $\tilde{D}$ of $\sigma$-algebra is considered for $\tilde{\mathcal{Y}}$, which is a block consisting of i.i.d. $\{\tilde{y_i}\}_{i=1}^{n}$ has the same length of $\mathcal{Y}$, we define the Rademacher complexity under this condition as $\mathfrak{R}^{n, \tilde{D}}_{SPO}(\mathcal{H})$.

\subsection{Generalization Bounds for the Dependent Data}

The following theorem is an application of the classical generalization bounds based on Rademacher complexity due to our setting.

\begin{corollary}(McDonald, Shalizi and Schervish \cite{mcdonald2017nonparametric})
For the $f \in \mathcal{H}$ has fixed memory l, we have that $f(y_{1:j}) = f(y_{j-l+1:j})$. Thus we have $R_{SPO, n}(f) = R_{SPO, l}(f)$. 
\label{coro. McDonald}
\end{corollary}

\begin{theorem}
\label{theorem. SPO_generalization_bound}(SPO Bounded Loss)
Let $\mathcal{H}$ be the class of functions mapping from $\mathbb{R}^{d\times l}$ to $\mathbb{R}^{d}$. Then, for any $m, a, l > 0$ with $2ma +l = n$, $\delta_{1} > 2m \beta(a-l)$, and $\delta_{2} > 4m \beta(a-l)$, with probability at least $1 - \delta_{1}$, the following inequality holds for all hypotheses $f \in \mathcal{H}$: 
\begin{equation}
R_{\mathrm{SPO},n}(f) \leq \widehat{R}_{\mathrm{SPO}, n}(f)+2 \mathfrak{R}_{\mathrm{SPO}}^{m, \tilde{D}}(\mathcal{H})+\omega_{S}(\mathfrak{Y}) \sqrt{\frac{\log (2 / \delta_{1}')}{2 m}},
\end{equation}
and with probability at least $1-\delta_{2}$, the following inequality holds for all hypotheses $f \in \mathcal{H}$:
\begin{equation}
R_{\mathrm{SPO},n}(f) \leq \widehat{R}_{\mathrm{SPO, n}}(f)+2 \widehat{\mathfrak{R}}_{\mathrm{SPO}, \tilde{\mathcal{Y}}}(\mathcal{H})+3 \omega_{S}(\mathfrak{Y}) \sqrt{\frac{\log (4 / \delta_{2}')}{2 m}},
\end{equation}
\end{theorem}

where $\delta_{1}^{\prime} = \delta_{1} - 2m \beta(a-l), \delta_{2}^{\prime} = \delta_{2} - 2m \beta(a-l)$ and the Rademacher Complexity characterized the complexity of the hypothesis class and the last term of inequality is bounded according to the value of the mixing coefficient $\beta(a-l)$, which gives a tighter bound than using the growth function and VC dimensions to quantify the generalization bound for the time series forecasting \cite{mcdonald2017nonparametric}.

\subsection{Calibration with Different Feasible Regions}
Adopting the definition in \cite{liu2021risk, ho2022risk,steinwart2007compare}, we use the calibration to characterize the relationship between the risk bound of a surrogate loss function $\ell$ and the risk bound of the SPO loss function.

\textbf{Polyhedral Sets}. When the feasible region of the decision vector is a polyhedron, we can derive a generalization bound using the combinatorial parameters, e.g. the extreme points of the feasible region, which are used to measure the function classes' complexity. 

\begin{theorem}(Calibration for Polyhedral Sets) Suppose that the feasible region $S$ is a bounded polyhedron, the optimal predictor and $f^{*}$ is in the hypothesis class $\mathcal{H}$, and there exists a constant $C'$ such that $\mathfrak{R}_{SPO}^{m,\tilde{D}}(\mathcal{H}) \leq \frac{C'}{\sqrt{m}}$. Let $\hat{f}_{SPO+}^{m}$ denote the predictor minimize the empirical SPO+ risk $\hat{R}_{SPO+}^{n}(\cdot)$ over $\mathcal{H}$. Then there exists a constant $C$ such that for any $\mathbb{P} \in \mathcal{P}_{Gaussian}$ and $\delta \in (0, \frac{1}{2})$, with probability at least $1 - \delta$, it holds that
\begin{equation}
\begin{aligned}
R_{SPO}(\hat{f}_{SPO+}^{m};
\mathbb{P}) - R_{SPO}^{*}(\mathbb{P}) \leq \frac{C \sqrt{\log(1/\delta')}}{m^{1/4}}.
\end{aligned}
\end{equation}
where $\delta^{\prime} = \delta - 2m\beta(a-l)$ and $R_{SPO}^{*}(\mathbb{P})$ is the minimum of the SPO generalization error. 
\label{theo. poly}
\end{theorem}

\textbf{Strongly Convex Level Sets}. When the feasible region of the decision vector is a strongly convex level set, i.e. there exists a $\mu$-strongly convex and $L$-smooth function $f: \mathbb{R}^{d} \rightarrow \mathbb{R}$ and the feasible region $S$ is bounded by $\{w \in \mathbb{R}^{d}| g(w) \leq r\}$ where $r$ is a constant that satisfies $r>f_{min}:= min_{w}f(w)$. 

\begin{theorem}(Calibration for Strongly Convex Level Sets) Suppose that the feasible region $S$ is a bounded polyhedron, the optimal predictor and $h^{*} = \mathbb{E}[\mathbf{y_{N+1}}|y_{1:N}]$ is in the hypothesis class $\mathcal{H}$, and there exists a constant $C'$ such that $\mathfrak{R}^{m}(\mathcal{H}) \leq \frac{C'}{\sqrt{m}}$. Let $\hat{h}_{SPO+}^{n}$ denote the predictor minimize the empirical SPO+ risk $\hat{R}_{SPO+}^{n}(\cdot)$ over $\mathcal{H}$. Then there exists a constant $C$ such that for any $\mathbb{P} \in \mathcal{P}_{Gaussian}$ and $\delta \in (0, \frac{1}{2})$, with probability at least $1 - \delta$, it holds that 
\begin{equation}
\begin{aligned}
R_{SPO}(\hat{g}_{SPO+}^{m};
\mathbb{P}) - R_{SPO}^{*}(\mathbb{P}) \leq \frac{C \sqrt{\log(1/\delta')}}{m^{1/2}}.
\end{aligned}
\end{equation}
\label{theo. strong}
\end{theorem}

where $\delta^{\prime} = \delta - 2m\beta(a-l)$ and $R_{SPO}^{*}(\mathbb{P})$ is the minimum of the SPO generalization error.

\section{\scshape Computational Experiments}\label{Computational}
\textbf{Knapsack Problem with Dynamic Cost Vectors}. In this section, considering a Knapsack Problem with cost vectors $\mathcal{Y}$ are observed from some random dynamic systems, where the cost vectors' trajectory $\mathcal{Y}$ is generated by an unknown stochastic dynamical system $\Phi$ as follows:

\begin{equation}
\label{eq.2}
    \Phi:\left\{\begin{array}{lll}
        x_{k+1} & = A x_{k}+\omega_{k}, & \omega_{k} \sim \mathcal{N}(0, Q), \\
        y_{k} & = o(x_{k}, \xi_{k}),
        \end{array}\right.
\end{equation}
where $x_{k}, y_{k} \in \mathbb{R}^{2}, A, Q\in \mathbb{R}^{2\times 2}$. $\{x_{k}\}_{k=1}^{+\infty}$ and $\{y_{k}\}_{k=1}^{+\infty}$ are the state vectors and cost vectors respectively. $o(\cdot,\cdot)$ is the observer and $\{\xi_{k}\}_{k=1}^{+\infty}$ is the observation noise. Without the loss of generality, we assume that the spectral radius of the state transition matrix $A$ is no larger than one. Thus, the nominal optimization problem can be formulated as:
\begin{equation}
\begin{array}{l}
\min \limits_{w \in S} \frac{1}{n} \sum\limits_{i=1}^{n} y_{i}^{\mathrm{T}}w_{k},\\
\text { s.t. } \Phi. \\
\end{array}
\end{equation}

In our experiment setting, the dimension of the cost vector d = 2 corresponds to 2 dimension vectors,  and $o(x_{k}, \xi_{k}) = ((H x_{k})^{deg} + 0.5*\mathbf{1})*\xi_{k}, k \in \mathbb{N}_{+}$, where $\mathbf{1}$ is a two-dimensional 1 vector. Firstly, we randomly generate different trajectories with different lengths of $q+p, q,p \in \mathbb{N}_{+}$ as the synthetic data set and $\xi_{k}, k \in \mathbb{N}_{+}$ is a multiplicative noise correlated to the observer of the cost vector and is generated form a uniform distribution $[1 - \overline{\xi}, 1+\overline{\xi}]$ with a noise half width parameter $\overline{\xi}$.

\textbf{Normalized Regret among Different 
Loss Functions}. This section presents experimental results for the regret by using synthetic datasets generated by Equation \ref{eq.2}. To compare the prediction performance of different loss functions, we fix the dynamics by setting $A = \begin{bmatrix}
            0.8 & 0.5\\
            0 & 0.8
            \end{bmatrix},$ $Q = \begin{bmatrix}
            0.1 & 0\\
            0 & 0.1
            \end{bmatrix}$. In these experiments, we examine the normalized regret with varied training set sizes of $q = 1000$, and $5000$, which means that we use the first $q$ time steps of the trajectory to train with the two-stage optimization method by using the least squares loss function $\ell_{2}$ and the absolute loss function $\ell_{1}$ and the SPO-based autoregressive framework. We select $p = 300$ time steps after the first $q$ time steps in sample trajectories as the test set when the prediction model is trained. The normalized regret is defined by:
\begin{equation}
    \frac{\sum_{i=1}^{q} \ell_{\text {SPO}}\left(\widehat{f}(y_{i-l+p-1: i+p-1}), y_{i+p}\right)}{\sum_{i=1}^{q}\left|z(y_{i+p})^{*}\right|},
\end{equation}

where $\widehat{f}$ is the trained prediction model, i.e. the predicted matrix $\widehat{M}$ and $z(y_{i})^{*}:= min_{w_{i} \in S} y_{i}^{\mathrm{T}}w_{i}$ is the true optimal cost of the test set. For all loss functions, we employ the Adam optimization method developed by \cite{kingma2014adam} and the PyEPO toolbox developed by \cite{tang2022pyepo} to train the parameters of the prediction models. It is important to note that loss functions $\ell_{2}$ and $\ell_{1}$ do not leverage the structure of the feasible region S and can thus be considered as purely learning the relationship between cost and feature vectors. In all of our settings, we run 1000 independent trials, i.e. 1000 independent trajectories, for each $q$ and picture the boxplots of the normalized SPO loss as follows.

\begin{figure}[htbp]
\flushleft
\includegraphics[width=14cm]{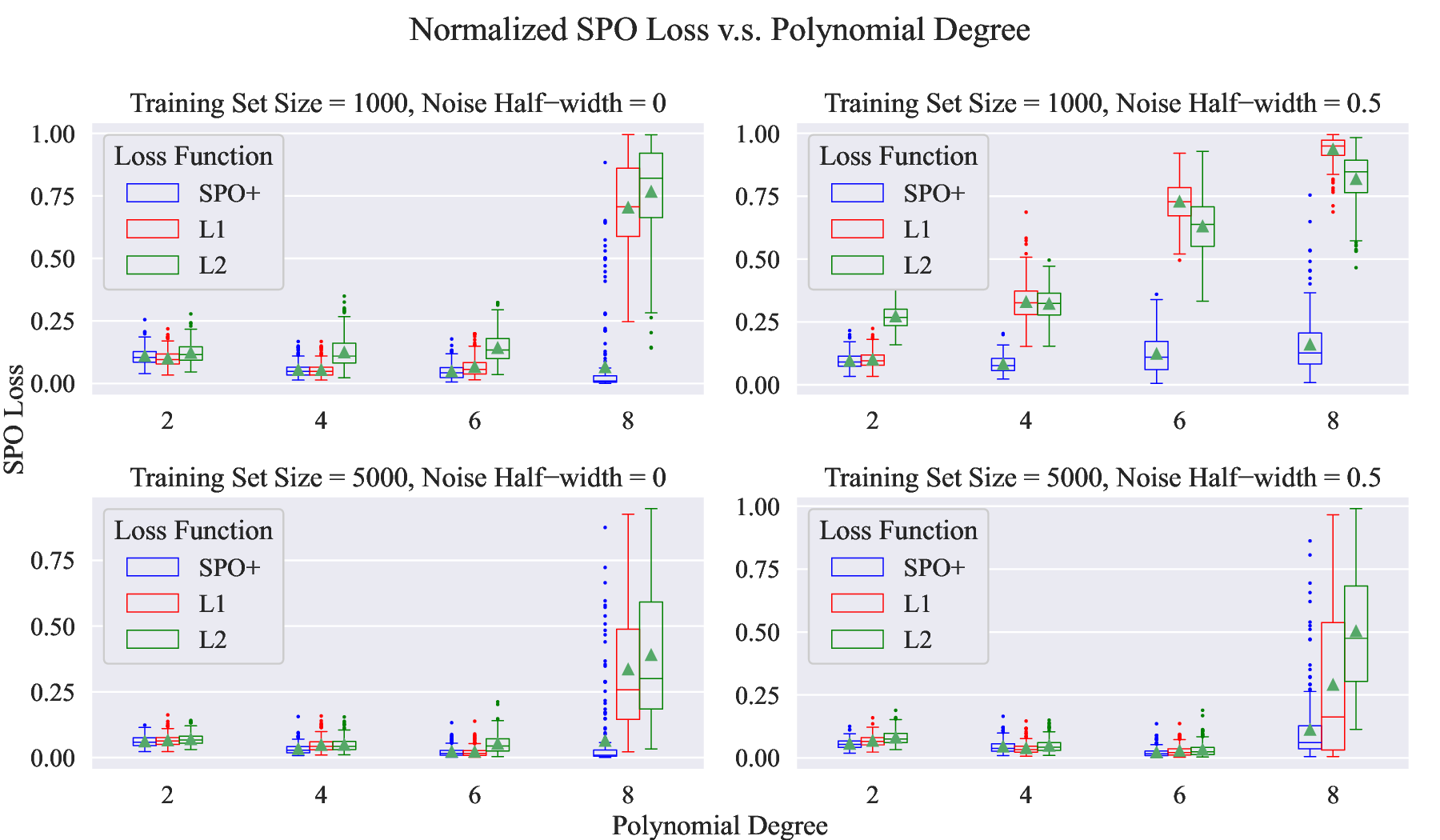}
\caption{Normalized test set SPO loss for the SPO+, least squares, and absolute loss methods on Knapsack instances.}
\label{fig. Normal}
\end{figure}

Figure \ref{fig. Normal} presents the empirical results for different loss functions. For smaller values of the degree parameters (i.e., $deg = 2$), the performances of all three methods are comparable. However, as the degree parameters increase, the SPO+ method outperforms the others. The superior performance of the $\ell_1$ loss compared to the squared $\ell_2$ loss can be attributed to its robustness against outliers in nonlinear dynamic systems. When $deg$ increases from 6 to 8, there is a significant rise in the $\ell_{1}$ and $\ell_{2}$ losses, which is due to the influence of noise in the system dynamics being nonlinearly amplified as the number of iterations increases. In other words, the influence of noise can either be negligible or extremely large, causing the regret to approach 1. Overall, Figure \ref{fig. Normal} demonstrates that the SPO+ loss excels in utilizing additional data and capturing stronger nonlinear signals compared to its competitors.

\textbf{Nomalized Regret with Different Mixing}. Besides, the mixing coefficients vary with different spectral radius of the matrix $A$, so we consider different state transition matrices with different spectral radius. To discover the impact of the mixing coefficient's value on the normalized regret, we first fixed $deg = 8$ of the observer. Similarly, the training set sizes $q = 1000$, and $5000$ with $Q = \begin{bmatrix}
            0.1 & 0\\
            0 & 0.1
            \end{bmatrix}$. The matrix $A = \begin{bmatrix}
            0.8 & a_{12}\\
            0 & 0.8
            \end{bmatrix}$, $p = 300$ for the test data set, and 1000 independent trials to picture the normalized SPO loss versus spectral radius. The experiments are carried out with different $a_{12} = 0, 0.1, 0.2, \ldots, 0.6$ and the spectral radius is increasing from $0.8$ to $1$ with $a_{12}$ increasing, i.e. system $\Phi$ gradually becomes marginal stable and $\beta(k)$ gradually becomes larger with fixed $k \in \mathbb{N}_{+}$.

\begin{figure}[htbp]
\flushleft
\includegraphics[width=14cm]{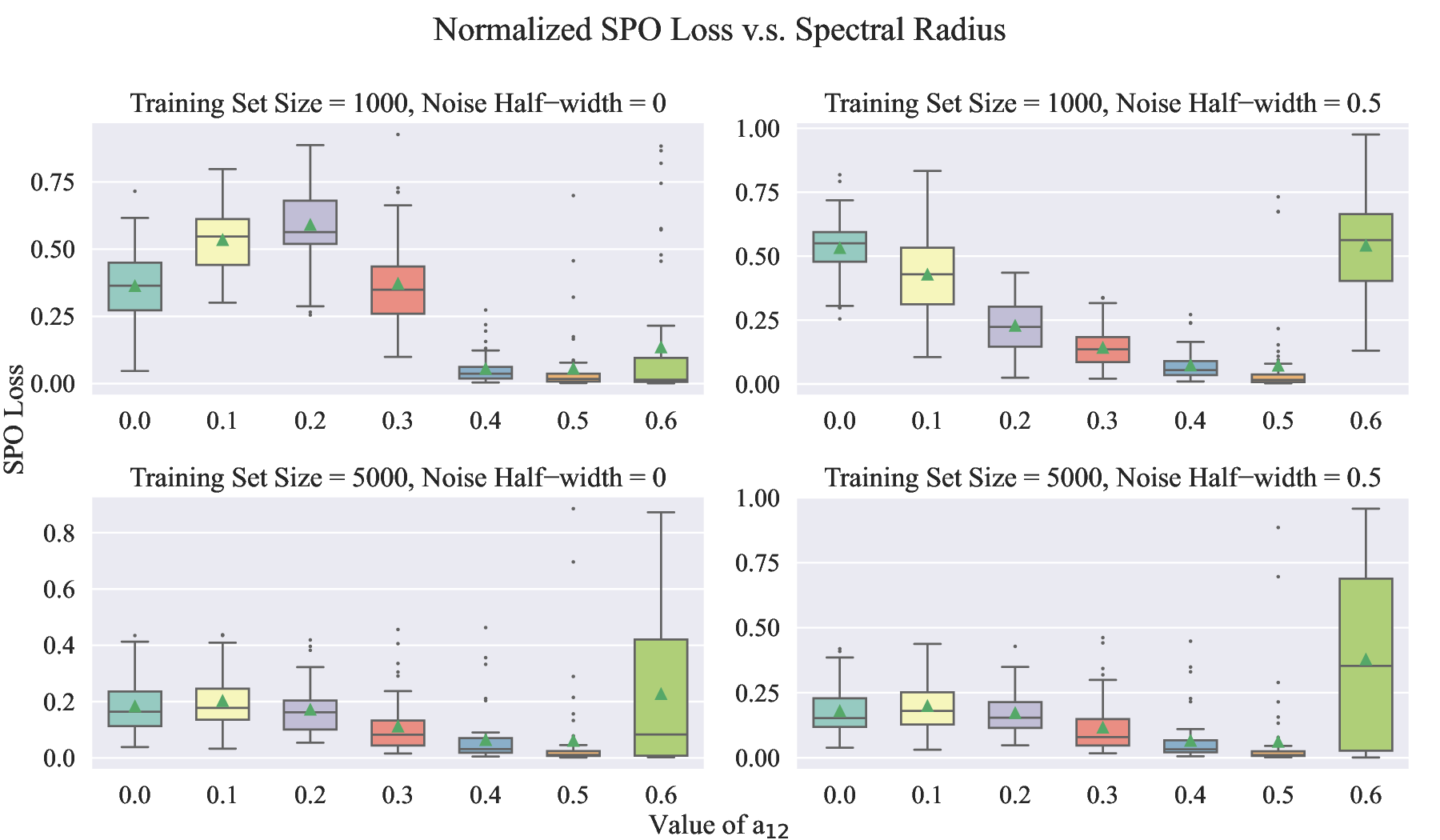}
\caption{Boxplot of the test set's normalized SPO loss for the SPO+ loss methods with enhancing mixing.}
\label{fig. Spectral}
\end{figure}

Here we focus on the relationship between the empirical loss and the spectral norm, i.e. the mixing coefficient $\beta(k)$ for $k \in \mathbb{N}_{+}$. As shown in Figure \ref{fig. Spectral}, when $a_{12} < 0.6$, different $a_{12}$, the empirical normalized SPO loss firstly increases slightly and then decreases to a relatively small value when the spectral radius becomes close to 1. When $a_{12} = 0.6$, i.e. the spectral norm is equal to 1, the empirical loss rapidly increases. In other words, the influence of noise can either be ignored, or it is extremely large so that the regret is close to 1. Generally speaking, Figure \ref{fig. Spectral} shows that the SPO+ loss is better than the competitors at leveraging additional data and stronger nonlinear signals for the sake of explaining the results revealed in Figure \ref{fig. Spectral}, we now need to understand the mixing time for a stable stationary process generated by a linear time-invariant (LTI) dynamical system. When the system satisfies $\Phi$ and the spectral radius $\rho$ is smaller than 1, then the system mixes at a linear rate, with the mixing coefficients $\beta(k) = o(\rho^{k})$ \cite{mokkadem1988mixing}. When $\rho$ becomes larger, the mixing coefficient $\beta(k)$ decays slower with the same $k$, and more information about the dynamics can be learned with the same lag $l$ in the autoregression model. Nevertheless, when $\rho$ becomes larger by no less than 1 and the system becomes marginally stable or unstable, the convergence rate shows the same degradation characteristic as in  \cite{simchowitz2018learning}, which will result in a sudden increase in the normalized SPO loss for the downstream optimization problem.

\section{\scshape Conclusions and Future Directions}\label{Conclusions}
Our work develops the learning theory of the online PTO problem to the circumstances when the data is obtained from a stationary mixing process: (1) We analyze the generalization bound with dependent data for the SPO loss using blocking techniques and the mixing coefficients which are used to quantify the dependency among entries of time series. (2) We build the uniform calibration between the SPO+ loss and the SPO loss for the time series. (3) We carry the experiments to prove the prevalence of the SPO framework even when the data is dependent. Nevertheless, it is important to note that the mixing time of a linear system deteriorates as the system becomes more unstable and ceases to exist for unstable systems. Consequently, techniques that do not use mixing time arguments in this study might be promising, especially by employing the prior knowledge of the dynamical system \cite{simchowitz2018learning,tsiamis2020sample,chiuso2019system,ziemann2024sharp}, under these assumptions, a tighter generalization bound is a worthwhile direction to research and develop. Also, how to design an autoregressive SPO framework that can use the information from all the historical data rather than just use limited data in fixed memory is challenging.

\bibliographystyle{citation4RB} 
\newpage
\appendix
\section{Redemancher Complexity and the Generalization Bounds}\label{App. GeneBound}

\textbf{Proof of Collary \ref{coro. McDonald}}
\begin{proof}
According to the fact that the model has a fixed memory, i.e. the lag length $l$, we have $f(Y_{1:j}) = f(Y_{j-d+1:j})$. Thus we have 
\begin{equation}
\begin{aligned}
\begin{array}{l}
R_{SPO, n}(f)  = \mathbb{E}\left[ \ell_{SPO}(Y_{n+1} - f(Y_{1:n}))\right] \\
= \mathbb{E}\left[\ell_{SPO}(Y_{n+1} - f(Y_{n-l+1:n}))\right]= \mathbb{E}\left[ \ell_{SPO}(Y_{l+1}) - f(Y_{1:l})\right] = R_{SPO, l}(f).
\end{array}
\end{aligned}
\end{equation}
\end{proof}

To prove Theorem \ref{theorem. SPO_generalization_bound}, we first state three lemmas to characterize the relationship between the generalization error and the empirical error.
\begin{lemma}(Lemma 4.1 in \cite{yu1994rates})
    Let $y$ be an event concerning the block sequence $\mathcal{Y}_{0}$. Then, we have
    \begin{equation}
        \left|\mathbb{P}(Z)-\tilde{\mathbb{P}}(Z)\right| \leq (m-1)\beta(a-l)
    \end{equation}
    \label{lemm. yu1994}
\end{lemma}

This lemma provides a method to apply results derived for i.i.d. data to $\beta$-mixing data. As the dependence between data points diminishes with increased separation between blocks, widely spaced blocks become nearly independent. Specifically, the discrepancy between expectations over these nearly independent blocks and those over truly independent blocks can be bounded by the $\beta$-mixing coefficient.

\begin{lemma}
    We denote the supreme of gap between the generalization error $R_{SPO,n}(f)$ and the empirical error $\hat{R}_{SPO,n}(f)$ for a given trajectory $\mathcal{Y}$ as $\psi(\mathcal{Y}, n) := \sup \limits_{f \in \mathcal{H}} [R_{SPO, n}(f) - \hat{R}_{SPO, n}(f)]$. And let $\epsilon' = \epsilon - \mathbb{E}_{\tilde{\mathcal{Y}}_{0}}[\psi(\tilde{\mathcal{Y}}_{0}, d)]$. Thus we have
    \begin{equation}
            \mathbb{P}_{\mathcal{Y}} \left( \psi(\mathcal{Y},n) >\epsilon \right) \leq 2\mathbb{P}_{\tilde{\mathcal{Y}_{0}}}\left(\psi(\tilde{\mathcal{Y}_{0}}, d) - \mathbb{E}_{\tilde{\mathcal{Y}_{0}}}\left[\psi(\tilde{\mathcal{Y}_{0}}, d)\right] > \epsilon \right) + 2 m \beta(a-l).
    \end{equation}
    \label{lemm. pro_gr}
\end{lemma}

\textbf{Proof of Lemma \ref{lemm. pro_gr}}.
\begin{proof}
According to the definition, we have
\begin{equation}
\begin{aligned}
    \begin{array}{l}
\mathbb{P}_{\mathcal{Y}}\left(\psi(\mathcal{Y}, n)>\epsilon\right) = \mathbb{P}_{\mathcal{Y}}\left(\psi(\mathcal{Y}, l)>\epsilon\right)\\\
=\mathbb{P}_{\mathcal{Y}}\left(\sup_{f \in \mathcal{H}} R_{l}(f)-\widehat{R}_{\mathcal{Y}}(f)>\epsilon\right)\\
=\mathbb{P}_{\mathcal{Y}}\left(\frac{1}{2}\left(\sup_{f \in \mathcal{H}}[R_{SPO, l}(f)-\hat{R}_{\mathcal{Y}_{0}}(f)+ R_{SPO, l}(f)-\widehat{R}_{\mathbf{\mathcal{Y}_{1}}}(f)]>\epsilon \right)\right) \\
\leq \mathbb{P}_{\mathcal{Y}}\left(\sup_{f \in \mathcal{H}} [R_{SPO, l}(f)-\widehat{R}_{\mathbf{\mathcal{Y}_{0}}}(f)]+\sup _{f \in \mathcal{H}} [R_{l}(f)-\widehat{R}_{\mathbf{\mathcal{Y}_{1}}}(f)] > 2 \epsilon\right) \\
\leq \mathbb{P}_{\mathcal{Y}_{0}}\left(\psi(\mathcal{Y}_{0},l)>\epsilon\right)+\mathbb{P}_{\mathcal{Y}_{1}}\left(\psi(\mathcal{Y}_{1},l) > \epsilon\right) \\
=2 \mathbb{P}_{\mathcal{Y}_{0}}\left(\psi(\mathcal{Y}_{0},l)>\epsilon\right) \\
=2 \mathbb{P}_{\mathcal{Y}_{0}}\left(\psi(\mathcal{Y}_{0},l) - \mathbb{E}_{\tilde{\mathcal{Y}}_{0}}[\psi(\tilde{\mathcal{Y}_{0}},l)]>\epsilon ' \right).
\end{array}
\end{aligned}
\end{equation}

The first equality is held by the fixed memory in Collary \ref{coro. McDonald}, the first inequality is held by the convexity of $\psi(\cdot, \cdot)$, and the second inequality is held by the union bound. According to the definition of $R_{l}(f)$ and the model we used, we knew that a prediction of in $\mathcal{Y}_{0}$ or $\mathcal{Y}_{1}$
depends on $l+1$ data values. Therefore, a prediction at any point in one block in $\mathcal{Y}_{0}$ is separated by at least $a-d$ observation from any different block in another different block. Therefore, apply Lemma \ref{lemm. yu1994}, we have:
\begin{equation}
    2\mathbb{P}_{\mathcal{Y}_{0}}\left(\psi(\mathcal{Y}_{0}, d) - \mathbb{E}_{\tilde{\mathcal{Y}}_{0}}[\psi(\tilde{\mathcal{Y}}_{0}, d)] > \epsilon \right) \leq 2\mathbb{P}_{\tilde{\mathcal{Y}_{0}}}\left(\psi(\tilde{\mathcal{Y}_{0}}, d) - \mathbb{E}_{\tilde{\mathcal{Y}_{0}}}[\psi(\tilde{\mathcal{Y}_{0}}, d)] > \epsilon \right) + 2 m \beta(a-l)
\end{equation}
\end{proof}

\begin{lemma}(Redemancher complexity bound \cite{mohri2008rademacher})
Let $H$ be a set of hypotheses bounded by $M \geq 0$. Then, for any sample $\mathcal{Y}$ of size n drawn from a stationary $\beta$-mixing distribution, and for any $m, a> 0$ with $2ma =n $ and $\delta_{1} > 2(m-1)\beta(a)$, $\delta_{2} > 4(m-1)\beta(a)$ with probability at least $1-\delta_{1}$, the following inequality holds for all hypotheses $h \in \mathcal{H}$: 
\begin{equation}
R(f) \leq \widehat{R}_{\mathcal{Y}}(f)+2\mathfrak{R}_{m}^{\widetilde{D}}(H)+M \sqrt{\frac{\log \frac{2}{\delta_{1}^{\prime}}}{2 m}}, 
\end{equation}
and with probability at least $1-\delta_{2}$, the following inequality holds for all hypotheses $f \in \mathcal{H}$:
\begin{equation}
R_(f) \leq \widehat{R}_{\mathcal{Y}}(f)+2 \widehat{\mathfrak{R}}_{m}^{\widetilde{D}}(\mathcal{H})+3M\sqrt{\frac{\log (\frac{4}{\delta_{2}^{\prime}})}{2 m}}.
\end{equation}
\label{lemm: reda}
\end{lemma}

where  $\delta_{1}^{\prime} = \delta_{1} - 2m \beta(a-l), \delta_{2}^{\prime} = \delta_{2} - 2m \beta(a-l)$.

\textbf{Proof of Theorem \ref{theorem. SPO_generalization_bound}}. 

\begin{proof}
According to Lemma \ref{lemm. pro_gr}, Lemma \ref{lemm. yu1994}, and Lemma \ref{lemm: reda}, and the definition of $\omega(\mathfrak{Y})$.
we have:
\begin{equation}
R_{\mathrm{SPO},n}(f) \leq \widehat{R}_{\mathrm{SPO}, n}(f)+2 \mathfrak{R}_{\mathrm{SPO}}^{m, \tilde{D}}(\mathcal{H})+\omega_{S}(\mathfrak{Y}) \sqrt{\frac{\log (2 / \delta_{1}')}{2 m}},
\end{equation}
and 
\begin{equation}
R_{\mathrm{SPO},n}(f) \leq \widehat{R}_{\mathrm{SPO, n}}(f)+2 \widehat{\mathfrak{R}}_{\mathrm{SPO}, \tilde{\mathcal{Y}}_{0}}(\mathcal{H})+3 \omega_{S}(\mathfrak{Y}) \sqrt{\frac{\log (4 / \delta_{2}')}{2 m}},
\end{equation}
\end{proof}

\section{Calibration Results}\label{App. Cali}
\subsection{Calibration between SPO+ and SPO Loss with i.i.d. data  \cite{liu2021risk}}
To prove Theorem \ref{theo. poly} and Theorem \ref{theo. strong}, we first state the calibration between the SPO loss and the SPO+ loss with i.i.d. data as follows. 

Here we introduce some additional notations for ease of explanation. we denote the generic norm as $\Vert \cdot \Vert_{*}$ which is defined by $\Vert \cdot \Vert_{*}$ is the dual norm. And the $A$-norm by $\Vert w \Vert := \sqrt{w^{T}Aw}$. The diameter of the set $S \subseteq \mathbb{R}^{d}$ is denoted as $D_{S} := sup_{w,w' \in S}\Vert w-w' \Vert _{2}$ and $d_{S}:=\min _{v \in \mathbb{R}^{d}:\|v\|_{2}=1}\left\{\max _{w \in S} v^{T} w-\min _{w \in S} v^{T} w\right\}$

Then we have the calibration result of Gaussian distribution for the Polyhedral Sets as: 

\begin{theorem}(Calibration functions for Polyhedral Feasible Sets)
Suppose that the feasible region $S$ is a bounded polyhedron and define $\Xi_{S} := (1+\frac{2\sqrt{3}D_{S}}{d_{s}})^{1-d}$. Then the calibration function of the SPO+ loss satisfies 
\begin{equation}
    \hat{\delta}_{\ell_{\mathrm{SPO}+}}\left(\epsilon ; \mathcal{P}_{Gaussian}\right) \geq \frac{\alpha \Xi_{S}}{4 \sqrt{2 \pi} e^{3}} \cdot \min \left\{\frac{\epsilon^{2}}{D_{S}}, \epsilon\right\} \quad \text { for all } \epsilon>0 \text {. }
\end{equation}
\end{theorem}

the calibration result of the Gaussian distribution for the strongly convex set: 

\begin{theorem}(Calibration functions for Strongly Convex Sets)
Suppose that the feasible region $S$ is bounded by a strongly convex function, i.e. there exists a $\mu$-strongly convex and $L$-smooth function $f: \mathbb{R}^{d} \rightarrow \mathbb{R}$ and the feasible region $S$ is bounded by $\{w \in \mathbb{R}^{d}| g(w) \leq r\}$ where $r$ is a constant that satisfies $r>f_{min}:= min_{w}f(w)$. Then it holds that, for any $\epsilon > 0$, it holds that $\hat{\delta}_{\ell_{\mathrm{SPO}}}\left(\epsilon ; \mathcal{P}_{Gaussian}\right) \geq \frac{\alpha \mu^{9 / 2}}{4 L^{9 / 2}} \cdot \epsilon$

\end{theorem}

\textbf{Proof of Theorem \ref{theo. poly}}
\begin{proof}
    According to the Theorem \ref{theorem. SPO_generalization_bound}, we have  
    \begin{equation}
        R_{\mathrm{SPO+},n}(f) - \widehat{R}_{\mathrm{SPO+}, n}(f) \leq 2 \mathfrak{R}_{\mathrm{SPO+}}^{m, \tilde{D}}(\mathcal{H})+\omega_{S}(\mathfrak{Y}) \sqrt{\frac{\log (2 / \delta_{1}')}{2 m}},
    \end{equation} 
    and the condition that $\mathfrak{R}_{SPO+}^{m,\tilde{D}}(\mathcal{H}) \leq \frac{C^{\prime}}{\sqrt{m}}$, then there exists some universal constant $C_{1}$ such that 
    \begin{equation}
        R_{\mathrm{SPO+},n}(f) - \widehat{R}_{\mathrm{SPO+}, n}(f) \leq C_{1}\sqrt{\frac{log(\frac{1}{\delta_{1}^{\prime}})}{m}}, 
    \end{equation}
    Besides, it is obvious that $\hat{R}_{SPO+,n}(\hat{f}_{SPO+}^{n})\leq \hat{R}_{SPO+,n}^{n}(f_{SPO+}^{*})$, thus with pr.obability at least $1-\delta$, we have 
    \begin{equation}
        R_{\mathrm{SPO+},n}(\hat{f}_{SPO}^{n}) - R_{\mathrm{SPO+}}^{*} \leq 2C_{1}\sqrt{\frac{log(\frac{1}{\delta_{1}^{\prime}})}{m}}
    \end{equation}
    According to the calibration results of Theorem \ref{theo. poly}, we have: 
    \begin{equation}
        R_{SPO}(\hat{f}_{SPO+}^{m}; \mathbb{P}) - R_{SPO}^{*}(\mathbb{P}) \leq \frac{C \sqrt{\log(1/\delta')}}{m^{1/4}}.
    \end{equation}
    where $\delta^{\prime} = \delta - 2m \beta(a-l)$
\end{proof}

\textbf{Proof of Theorem \ref{theo. strong}}
\begin{proof}
    Similarly to the proof of the Theorem \ref{theo. poly}, we have 
    \begin{equation}
    R_{SPO}(\hat{g}_{SPO+}^{m};
    \mathbb{P}) - R_{SPO}^{*}(\mathbb{P}) \leq \frac{C \sqrt{\log(1/\delta')}}{m^{1/2}}.
    \end{equation}
    where $\delta^{\prime} = \delta - 2m \beta(a-l)$.
\end{proof}

\section{Discussion of the Limitations}\label{App. limit}
The blocking techniques used in the stationary and $\beta$-mixing process will result in a conventional generalization bound due to the deflating of the training data, which is unavoidable in the worst-case scenarios as shown by Nagaraj et al \cite{nagaraj2020least}. Besides, for realizable linear regression problems, mixing time impacts only the burn-in period and not the final risk. However, this conclusion relies on specific modifications of Stochastic Gradient Descent (SGD) and is not broadly applicable across different algorithms.

While lack of ergodicity does not affect parameter recovery rates in linear system identification, the same may not hold true for more complex dynamical systems, especially those driven by generalized linear models (GLMs). However, some studies indicate that ergodicity does not necessarily hamper parameter recovery rates even in semiparametric settings with unknown link functions. New techniques are needed for more complex dynamical systems.

\section{Experiments Details}\label{App. Exp}
For both experiments, we ran each instance on 8 cores of Intel Core i7-13700H @ 2.4 GHz.
\end{document}